# Stability and Robustness of the Disturbance Observer-based Motion Control Systems in Discrete-Time Domain

Emre Sariyildiz, *Member, IEEE*, Satoshi Hangai, *Member, IEEE*, Tarik Uzunovic, *Senior Member, IEEE*, Takahiro Nozaki, *Member, IEEE*, Kouhei Ohnishi, *Life Fellow, IEEE*

*Abstract*— This paper analyses the robust stability and performance of the Disturbance Observer- (DOb-) based digital motion control systems in discrete-time domain. It is shown that the phase margin and the robustness of the digital motion controller can be directly adjusted by tuning the nominal plant model and the bandwidth of the observer. However, they have upper and lower bounds due to robust stability and performance constraints as well as noise-sensitivity. The constraints on the design parameters of the DOb change when the digital motion controller is synthesised by measuring different states of a servo system. For example, the bandwidth of the DOb is limited by noise-sensitivity and *waterbed effect* when velocity and position measurements are employed in the digital robust motion controller synthesis. The robustness constraint due to the *waterbed effect* is removed when the DOb is implemented by acceleration measurement. The design constraints on the nominal plant model and the bandwidth of the observer are analytically derived by employing the generalised Bode Integral Theorem in discrete-time. The proposed design constraints allow one to systematically synthesise a high-performance DOb-based digital robust motion controller. Experimental results are given to verify the proposed analysis and synthesis methods.

## I. INTRODUCTION

TO SUPPRESS internal and external disturbances, observer-based robust controllers have been applied to many different motion control applications [1–5]. Among them, the DOb has become one of the most widely used robust motion control tools thanks to its simplicity and efficacy in practice [6–8]. Control engineering practitioners have applied this robust motion control tool to various advanced engineering systems by using intuitive design methods in which the control parameters of the DOb (e.g., the nominal model of the servo system and the bandwidth of the observer) are tuned by trial and error [7, 10, 11]. However, the stability and performance of the robust motion controller highly depend on designers' own experience when the DOb is intuitively synthesised. Several studies have been conducted to improve the stability and performance of the DOb-based robust motion control systems [1, 7, 12]. Nevertheless, there still lacks practical analysis and synthesis techniques in which the design parameters of the robust motion controller are tuned systematically [7, 13].

The successful engineering implementations have motivated many control engineering theoreticians to build new analysis and synthesis methods for the DOb-based robust motion control systems in the last two decades [7, 12]. Although the robust motion controllers are usually implemented by using computers or microcontrollers, continuous-time analysis and synthesis methods are generally employed due to simplicity [6, 8, 14, 15]. Continuous-time analysis methods, however, fall-short when explaining some dynamic behaviours of the DOb-based robust digital motion controllers. For example, as shown in this paper, the robust stability constraint on the bandwidth of the DOb becomes more severe due to discrete-time implementations. This may cause some unexpected dynamic behaviours such as poor stability and performance as the bandwidth of the DOb is increased [16–18]. To tackle this problem, discrete-time analysis and synthesis methods have been proposed for the DOb-based robust motion control systems. Lee et al. and Endo et al. used bilinear transformation to synthesise a DOb in discrete-time [19, 20]. Tesfaye et al. designed a DOb by using a sensitivity optimization method that provides better tracking performance than bilinear transformation in digital motion control applications [21]. Fujimoto et al. proposed a multi-rate sampling control method to improve the performance of trajectory tracking and disturbance suppression of servo systems [22]. Godler et al. and Bertoluzzo et al. showed that the DOb-based robust motion control systems may exhibit under-damped and even unstable responses as the bandwidth of disturbance estimation increases [17, 18]. Chen et al. tuned the dynamics of nominal plant model by using an optimal plant shaping method so as to increase the bandwidth of the DOb

E. Sariyildiz is with the School of Mechanical, Materials, Mechatronic and Biomedical Engineering, University of Wollongong, Wollongong, NSW, 2522, Australia. (e-mail: emre@uow.edu.au).

S. Hangai, and K. Ohnishi are with the Department of System Design Engineering, Keio University, Yokohama 223-8522, Japan. (e-mail: hangai@sum.sd.keio.ac.jp; ohnishi@sd.keio.ac.jp)

T. Uzunovic is with the Department of Automatic Control and Electronics, Faculty of Electrical Engineering, University of Sarajevo, Sarajevo 71000, Bosnia and Herzegovina (e-mail: tuzunovic@etf.unsa.ba).

T. Nozaki is with the Department of System Design Engineering, Keio University, Yokohama 223-8522, Japan and with the Department of Mechanical Engineering, Massachusetts Institute of Technology, Cambridge, MA 02139, USA (e-mail: nozaki@sd.keio.ac.jp).

[23]. Kong et al. optimised the parameters of the discretised nominal plant model to enhance the robust stability of the digital motion controller [24]. Antonello and Oboe combined position and acceleration measurements by using Kalman filter to increase the bandwidth of disturbance estimation [25]. Some advanced robust motion controllers were recently proposed by designing the DOb in discrete-time [26–28]. Despite the studies conducted in the literature, the stability and robustness constraints of the DOb-based motion control systems have not yet been derived in discrete-time. Moreover, although it is a well-known fact that measurement methods may significantly change the trade-off between the performance of disturbance estimation and noise-sensitivity [29], the influence of the measured state on the robust stability of the DOb-based motion control systems has not yet been discussed in the literature.

In this paper, the DOb-based robust motion control systems are analysed and synthesised in discrete-time domain. By employing Bode Integral Theorem, it is shown that continuous-time analysis methods fall-short in explaining some dynamic behaviours, such as poor stability and performance, of the DOb implemented by computers. The main reason is that the design constraints on the nominal plant model and the bandwidth of the observer become more severe due to digital implementation. This paper, for the first time, derives the stability and robustness constraints of the DOb in discrete-time. The proposed discrete-time analysis shows that the nominal plant model (i.e., phase margin) and the bandwidth of the observer (i.e., robustness) are limited not only by noise-sensitivity but also by *waterbed effect* when the DOb is synthesised by using velocity or position measurement. In order to eliminate the design constraint due to *waterbed effect*, we propose an acceleration measurement-based DOb. New stability and robustness constraints on the design parameters of the DOb are analytically derived. The proposed design constraints not only explain the stability and robustness of the DOb implemented by computers but also facilitate the digital robust motion controller synthesis. Simulation and experimental results are given to validate the proposed analysis and synthesis methods.

Within this context, the chief contributions of this paper are i) theoretically and experimentally proving that continuous-time analysis methods fall-short in explaining the robust stability and performance of the DOb-based digital robust motion control systems ii) deriving the design constraints on the nominal plant model and the bandwidth of the observer implemented by computers/microcontrollers iii) showing that not only the noise sensitivity but also the robust stability and performance of the DOb may significantly change by measured state (i.e., position, velocity and acceleration measurement) and iv) proposing new practical tools to systematically tune the design parameters of the DOb-based digital robust motion controllers.

The paper is organised as follows. In Section II, the limitations of continuous-time stability and robustness analysis methods are presented. In Section III, acceleration, velocity and position (pseudo-velocity) measurement-based DObs are analysed in discrete-time and the design constraints of the robust motion controllers are analytically derived. In Section IV, the stability and robustness of the DOb-based digital motion control systems are analysed. In Section V, the proposed analysis and synthesis methods are verified by experiments. In Section VI, the paper is concluded.

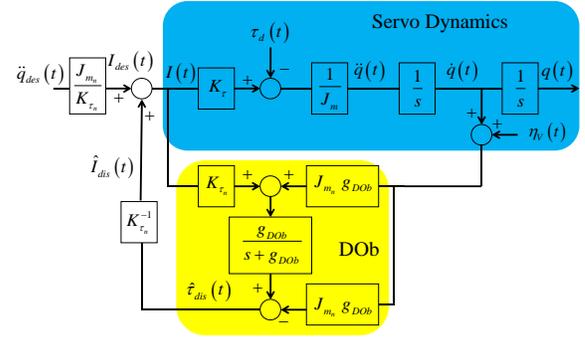

Fig. 1: Block diagram of the conventional DOb in continuous-time domain.

## II. CONTINUOUS-TIME ANALYSIS OF DOb

Figure 1 illustrates the block diagram of the conventional DOb (i.e., the inner-loop of the robust controller) in continuous-time domain [1, 6]. In this figure, $J_m$ and $J_{m_n}$ are the uncertain and nominal inertiae, respectively; $K_\tau$ and $K_{\tau_n}$ are the uncertain and nominal thrust coefficients, respectively; $\tau_d$ and $\eta_V$ are the disturbance and noise exogenous inputs, respectively; $q$, $\dot{q}$ and $\ddot{q}$ are the angle, velocity and acceleration of a servo system, respectively; $g_{DOb}$ is the bandwidth of the DOb; $I$ is the current of a DC motor; $\tau_{dis}$ and $I_{dis}$ are the internal and external disturbance torque and current, respectively; $\hat{\bullet}$ is the estimation of $\bullet$; $\bullet_{des}$ is the desired $\bullet$; and $s$ is complex Laplace variable [7].

The dynamic model of the servo system and disturbance estimation can be obtained from Fig. 1 as follows:

$$\ddot{q}(t) = J_m^{-1} K_\tau I(t) - J_m^{-1} \tau_d(t) \qquad (1)$$

$$\hat{\tau}_{dis}(s) = \left(K_{\tau_n} I(s) + J_{m_n} g_{DOb} \dot{q}_n(s)\right) \frac{g_{DOb}}{s + g_{DOb}} - J_{m_n} g_{DOb} \dot{q}_n(s) \qquad (2)$$

where $\dot{q}_n(t) = \dot{q}(t) + \eta_V(t)$ is the measured velocity state that includes noise; $I(t) = I_{des}(t) + \hat{I}_{dis}(t)$ and $\hat{\tau}_{dis}(t) = K_{\tau_n} \hat{I}_{dis}(t)$.

By substituting Eq. (2) into Eq. (1), the transfer functions of the inner-loop can be derived as follows:

$$\ddot{q}(s) = C^i(s)\ddot{q}_{des}(s) - J_m^{-1} S^i(s)\tau_d(s) - sT^i(s)\eta_V(s) \qquad (3)$$

where $S^i(s) = \dfrac{1}{1+L^i(s)} = \dfrac{s}{s+\alpha g_{DOb}}$ and $T^i(s) = \dfrac{L^i(s)}{1+L^i(s)} = \dfrac{\alpha g_{DOb}}{s+\alpha g_{DOb}}$ are the inner-loop's sensitivity and complementary sensitivity transfer functions in which $L^i(s) = \alpha g_{DOb}/s$ is the open-loop transfer function of the inner-loop and $\alpha = \left(J_{m_n} K_\tau\right)/\left(J_m K_{\tau_n}\right)$; and $C^i(s) = \alpha \dfrac{s+g_{DOb}}{s+\alpha g_{DOb}}$ is a phase-lead/lag compensator.

Equation 3 shows that the phase margin and robustness of the motion controller can be improved by increasing $\alpha$ and $g_{DOb}$, respectively. The inner-loop's transfer functions are stable for all values of the design parameters of $\alpha$ and $g_{DOb}$.

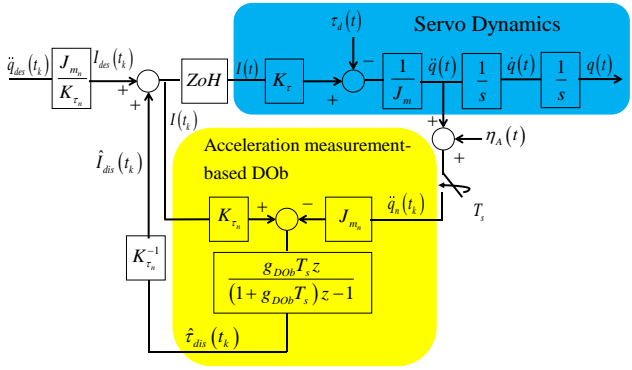

Fig. 2: Block diagram of the acceleration measurement-based DOb.

To analyse the robustness, let us apply the generalised Bode Integral Theorem to the conventional DOb-based motion controller illustrated in Fig. 1. The generalised Bode's integral equation is as follows:

$$\int_0^\infty \log\left(\left|S^i(j\omega)\right|\right)d\omega = \pi\sum_k \text{Re}\left(p_{u_k}^i\right) - \frac{\pi}{2}\lim_{s\to\infty} sL^i(s) = -\frac{\pi}{2}\alpha g_{DOb} \quad (4)$$

where $\omega$ is frequency, $j^2 = -1$ is complex number and $\text{Re}\left(p_{u_k}^i\right)$ is the real part of the $k^{th}$ right-half-plane pole of $L^i(s)$ [30, 31].

Since the right side of Eq. (4) has smaller values as $\alpha$ and $g_{DOb}$ are increased, the Bode's integral equation can hold without exhibiting a high sensitivity peak. In other words, Eq. (4) shows that the conventional DOb-based robust motion controller does not suffer from *waterbed effect* and good robust stability and performance can be achieved for all values of $\alpha$ and $g_{DOb}$. The reader is referred to [15, 30] for further details on the design constraints of the DOb in continuous-time domain.

However, this is not what we observe in the practical applications of the conventional DOb-based digital motion control systems. For example, the digital robust motion controller exhibits oscillatory response and becomes unstable as the bandwidth of the DOb increases in practice [17, 18]. Since continuous-time analysis methods fall-short in explaining the robust stability and performance of the DOb-based digital motion controllers, unexpected dynamic responses, such as poor stability, occur when the phase margin and robustness of the controller are improved by increasing $\alpha$ and $g_{DOb}$ [7, 30].

To explain the dynamic behaviour of the digital robust motion controller in practice, let us now analyse the stability and robustness of the DOb in discrete-time domain.

### III. DISCRETE-TIME ANALYSIS AND SYNTHESIS OF DOb

In this section, the stability and robustness of the DOb are analysed by considering acceleration, velocity and position measurements in discrete-time domain.

#### A. Acceleration Measurement-based DOb:

Block diagram of the acceleration measurement-based DOb is illustrated in Fig. 2. In this figure, $\eta_A$ is the noise exogenous input; $ZoH$ is the Zero-order Hold; $T_s$ is the sampling time; $t$ and $t_k = kT_s$ are time in continuous and discrete domains, respectively; $k = 1, 2, \ldots$ is an integer; and $z = e^{sT_s}$ is a complex variable. The other parameters are same as defined earlier.

By using Fig. 2, the dynamic model of the servo system can be described in discrete-time as follows:

$$\ddot{q}(t_k) = J_m^{-1}K_\tau I(t_k) - J_m^{-1}\tau_d(t_k) \quad (5)$$

$$\hat{\tau}_{dis}(z) = \left(K_{\tau_n}I(z) - J_{m_n}\ddot{q}_n(z)\right)\frac{g_{DOb}T_s z}{(1+g_{DOb}T_s)z - 1} \quad (6)$$

where $\ddot{q}_n(t_k) = \ddot{q}(t_k) + \eta_A(t_k)$ is the measured acceleration state; $I(t_k) = I_{des}(t_k) + \hat{I}_{dis}(t_k)$ and $\hat{\tau}_{dis}(t_k) = K_{\tau_n}\hat{I}_{dis}(t_k)$ at time $t_k = kT_s$.

The inner-loop transfer functions between the exogenous inputs and the acceleration-output are derived by substituting Eq. (6) into Eq. (5).

$$\ddot{q}(z) = C_A^i(z)\ddot{q}_{des}(z) - J_m^{-1}S_A^i(z)\tau_d(z) - T_A^i(z)\eta_A(z) \quad (7)$$

where $S_A^i(z) = \frac{1}{1+L_A^i(z)} = \frac{z-1}{(1+\alpha g_{DOb}T_s)z - 1}$ and $T_A^i(z) = \frac{L_A^i(z)}{1+L_A^i(z)}$
$= \frac{\alpha g_{DOb}T_s z}{(1+\alpha g_{DOb}T_s)z - 1}$ are the inner-loop's sensitivity and complementary sensitivity transfer functions in which $L_A^i(z) = \frac{\alpha g_{DOb}T_s z}{z-1}$ is the open-loop transfer function; and $C_A^i(z) = \alpha \frac{(1+g_{DOb}T_s)z - 1}{(1+\alpha g_{DOb}T_s)z - 1}$ is a phase-lead/lag compensator.

Equation (7) shows that the phase-lead (phase-lag) compensator $C_A^i(z)$ is synthesised in the inner-loop when the nominal inertia and thrust coefficient are properly tuned so that $\alpha > 1$ ($\alpha < 1$). The phase-margin of the robust motion controller improves by increasing $\alpha$, i.e., using larger (smaller) values of the nominal inertia (thrust coefficient). The inner-loop transfer functions are stable for all values of the nominal plant parameters and the bandwidth of the DOb. As $\alpha$ or $g_{DOb}$ increases, lower values for the frequency response of the sensitivity function are obtained. This improves the robustness of the motion controller against disturbances (see Fig. 3a).

Let us employ the generalised Bode Integral Theorem to analyse the robust stability and performance of the DOb-based digital robust motion controller. The Bode Integral equation of the motion controller illustrated in Fig. 2 is as follows:

$$\int_{-\pi}^{\pi} \ln\left|S_A^i\left(e^{j\omega T_s}\right)\right|d\omega T_s = -2\pi\ln\left|1+\psi_A(z)\right| = -2\pi\ln\left|1+\alpha g_{DOb}T_s\right| \quad (8)$$

where $\psi_A(z) = \lim_{z\to\infty} L_A^i(z) = \alpha g_{DOb}T_s$ [31].

The right side of Eq. (8) has smaller values as $\alpha$ and $g_{DOb}$ are increased. Therefore, the Bode integral equation can hold

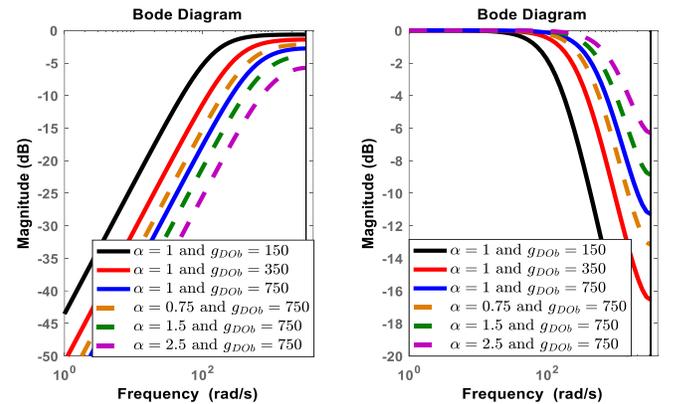

a) Sensitivity function; b) Complementary sensitivity function.
Fig. 3: Frequency responses of $S_A^i(z)$ and $T_A^i(z)$ when $T_s = 1$ms.

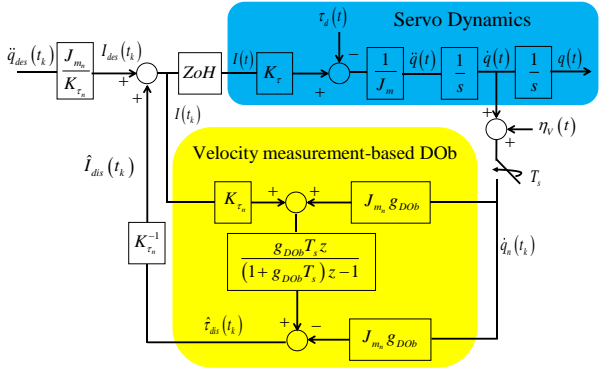

Fig. 4: Block diagram of the velocity measurement-based DOb.

without exhibiting high sensitivity peaks when the robustness against disturbances is improved at low frequencies. In other words, the acceleration measurement-based DOb is not influenced by *waterbed effect* as shown in Fig. 3 [15, 30]. Although disturbances are not excited by the sensitivity peaks, the noise of acceleration measurement is suppressed less as $\alpha$ and $g_{DOb}$ are increased (see Fig 3b). Therefore, the bandwidth of the DOb and the phase-margin of the digital robust motion controller are limited by only the noise-sensitivity of the acceleration measurement system.

### B. Velocity Measurement-based DOb:

Figure 4 illustrates the velocity measurement based- (i.e., conventional) DOb in discrete-time. Similarly, the following equations can be directly obtained from Fig. 4 for the discrete model of the servo system and disturbance estimation.

$$\dot{q}(z) = J_m^{-1} K_\tau \frac{T_s}{z-1} I(z) - J_m^{-1} \frac{T_s}{z-1} \tau_d(z) \quad (9)$$

$$\hat{\tau}_{dis}(z) = \left(K_{\tau_n} I(z) + J_{m_n} g_{DOb} \dot{q}_n(z)\right) \frac{g_{DOb} T_s z}{(1+g_{DOb} T_s) z - 1} - J_{m_n} g_{DOb} \dot{q}_n(z) \quad (10)$$

where $\dot{q}_n(t_k) = \dot{q}(t_k) + \eta_V(t_k)$ is the measured velocity at time $t_k$.

By substituting Eq. (10) into Eq. (9), the inner-loop transfer functions between the exogenous inputs and the velocity-output are derived as follows:

$$\dot{q}(z) = C_V^i(z) G_V(z) \ddot{q}_{des}(z) - J_m^{-1} S_V^i(z) G_V(z) \tau_d(z) - T_V^i(z) \eta_V(z) \quad (11)$$

where $S_V^i(z) = \frac{1}{1+L_V^i(z)} = \frac{z-1}{z-(1-\alpha g_{DOb} T_s)}$ and $T_V^i(z) = \frac{L_V^i(z)}{1+L_V^i(z)} = \frac{\alpha g_{DOb} T_s}{z-(1-\alpha g_{DOb} T_s)}$ are the inner-loop's sensitivity and complementary sensitivity transfer functions in which $L_V^i(z) = \frac{\alpha g_{DOb} T_s}{z-1}$; $G_V(z) = \frac{T_s}{z-1}$ is the discrete model of the servo system when velocity measurement is used in the robust controller synthesis; and $C_V^i(z) = \alpha \frac{(1+g_{DOb} T_s) z - 1}{z-(1-\alpha g_{DOb} T_s)}$ is a phase-lead/lag compensator.

Equation (11) shows that the phase-lead (phase-lag) compensator $C_V^i(z)$ is synthesised in the inner-loop when $\alpha$ and $g_{DOb}$ are tuned so that $\alpha > 1/(1+g_{DOb} T_s)$ ($\alpha < 1/(1+g_{DOb} T_s)$). The phase margin of the robust motion controller can be increased by using larger values for the design parameter $\alpha$.

Similarly, the robustness against disturbances can be improved by increasing $\alpha$ and/or $g_{DOb}$. However, neither $\alpha$ nor $g_{DOb}$ can be freely tuned due to the stability constraint of the velocity measurement-based DOb. The robust motion controller becomes unstable when $\alpha g_{DOb} > 2/T_s$. To achieve non-oscillatory response (i.e., assign the discrete pole of the inner-loop between 0 and 1), the design parameters of the velocity measurement-based DOb should satisfy the following constraint.

$$\alpha g_{DOb} < 1/T_s \quad (12)$$

When the Bode Integral Theorem is applied to the robust motion controller in Fig. 4, the following equation is obtained.

$$\int_{-\pi}^{\pi} \ln \left| S_V^i \left(e^{j\omega T_s}\right) \right| d\omega T_s = -2\pi \ln \left|1+\psi_V(z)\right| = 0 \quad (13)$$

where $\psi_V(z) = \lim_{z \to \infty} L_V^i(z) = 0$ [31].

Equations (11) and (13) show that as $\left|S_V^i(z)\right|$ is decreased (i.e., the robustness against disturbances is improved) at low frequencies by using higher values of $\alpha$ and $g_{DOb}$, the peak of the sensitivity function increases at middle/high frequencies to hold the Bode's integral equation. In other words, *waterbed effect* occurs when the design parameters of the velocity measurement-based DOb are not properly tuned. The frequency responses of the sensitivity and complementary sensitivity functions are illustrated in Fig. 5. It is clear from this figure that as $\alpha$ and $g_{DOb}$ are increased, the robustness against disturbances improves at low frequencies yet the digital motion controller becomes more sensitive to disturbances at middle/high frequencies. For example, the noise of velocity measurement is suppressed less and may even be excited by the peak of the complementary sensitivity function due to *waterbed effect* (see Fig. 5b). This also degrades the robust stability of the digital motion controller.

Compared to the acceleration measurement-based DOb, the design parameters of the velocity measurement-based DOb are constrained by the stability and *waterbed effect*. To conduct high-performance motion control applications, the stability and robustness constraints should be considered in the digital motion controller synthesis. The upper bounds on the plant-model mismatch and the bandwidth of the DOb can be relaxed by employing higher sampling frequency; however, this generally increases the cost in engineering applications.

It should be noted here that the stability and robustness constraints on the nominal plant model and the bandwidth of the velocity measurement-based (i.e. conventional) DOb cannot

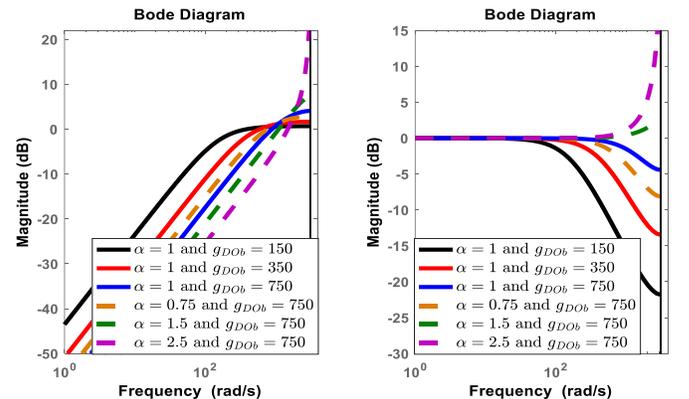

a) Sensitivity function; b) Complementary sensitivity function.
Fig. 5: Frequency responses of $S_V^i(z)$ and $T_V^i(z)$ when $T_s = 1$ms.

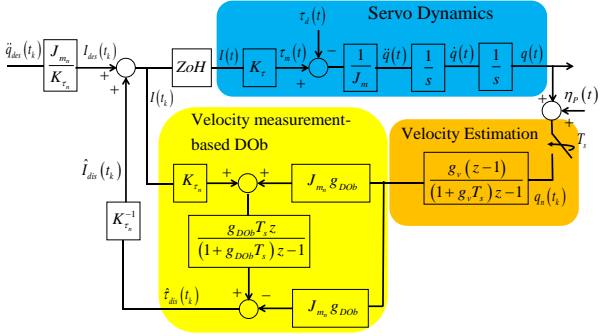

Fig. 6: Block diagram of the position measurement-based DOb.

be derived by employing continuous-time analysis methods. Although the digital robust motion controller may be influenced by *waterbed effect*, continuous-time analysis shows that good robust stability and performance can be achieved for all values of $\alpha$ and $g_{DOb}$ (see Section II). Thus, continuous-time analysis methods fall-short in explaining the dynamic behaviour of the DOb-based digital robust motion control systems.

### C. Position (Pseudo-Velocity) Measurement-based DOb:

The conventional DOb is generally implemented by differentiating position measurement (i.e., using pseudo-velocity measurement) as shown in Fig. 6. In this figure, $g_v$ is the bandwidth of velocity estimation and $\eta_P$ is the noise of position measurement. The other parameters are defined earlier.

When the DOb is implemented by measuring position state, the discrete model of the servo system and the estimated disturbance are as follows:

$$q(z) = J_m^{-1} K_\tau \frac{T_s^2(z+1)}{2(z-1)^2} I(z) - J_m^{-1} \frac{T_s^2(z+1)}{2(z-1)^2} \tau_d(z) \quad (14)$$

$$\hat{\tau}_{dis}(z) = \left( K_{\tau_n} I(z) + J_{m_n} g_{DOb} \frac{g_v(z-1)}{(1+g_v T_s)z-1} q_n(z) \right) \frac{g_{DOb} T_s z}{(1+g_{DOb} T_s)z-1} \\ - J_{m_n} g_{DOb} \frac{g_v(z-1)}{(1+g_v T_s)z-1} q_n(z) \quad (15)$$

where $q_n(t_k) = q(t_k) + \eta_P(t_k)$ is the measured position at time $t_k$.

The inner-loop transfer functions of the DOb can be similarly derived by substituting Eq. (15) into Eq. (14).

$$q(z) = C_P^i(z) G_P(z) \ddot{q}_{des}(z) - J_m^{-1} S_P^i(z) G_P(z) \tau_d(z) \\ - T_P^i(z) \eta_P(z) \quad (16)$$

where $S_P^i(z) = \frac{1}{1+L_P^i(z)} = \frac{(z-1)((1+g_v T_s)z-1)}{(z-1)((1+g_v T_s)z-1)+\beta g_v g_{DOb}(z+1)}$

and $T_P^i(z) = \frac{L_P^i(z)}{1+L_P^i(z)} = \frac{\beta g_v g_{DOb}(z+1)}{(z-1)((1+g_v T_s)z-1)+\beta g_v g_{DOb}(z+1)}$

are the inner-loop's sensitivity and complementary sensitivity transfer functions in which $L_P^i(z) = \frac{\beta g_v g_{DOb}(z+1)}{(z-1)((1+g_v T_s)z-1)}$ and $\beta = \frac{1}{2}\alpha T_s^2$; $G_P(z) = \frac{T_s^2}{2}\frac{z+1}{(z-1)^2}$ is the discrete model of the servo system when position measurement is employed; and

$$C_P^i(z) = \alpha \frac{\left((1+g_{DOb}T_s)z-1\right)\left((1+g_v T_s)z-1\right)}{(1+g_v T_s)z^2-(2+g_v T_s-\beta g_v g_{DOb})z+1+\beta g_v g_{DOb}}$$ is a phase-lead/lag compensator.

Equation (16) shows that the phase margin of the robust motion controller can be similarly improved by using higher values of $\alpha$ in the design of the position measurement-based DOb. Furthermore, the robustness against disturbances can be improved by increasing $\alpha$ and/or $g_{DOb}$. However, the design parameters of the DOb have upper bounds due to the stability constraint. When $\alpha g_{DOb} > 2/T_s$, the robust motion controller becomes unstable as the double poles of the inner-loop are out of the unit circle. To achieve non-oscillatory response by placing all discrete poles of the inner-loop between 0 and 1, the design parameters of the position measurement-based DOb should satisfy the following constraint.

$$\alpha g_{DOb} < \frac{6}{T_s} + \frac{8-\sqrt{32(2+g_v T_s)(1+g_v T_s)}}{g_v T_s^2} \quad (17)$$

When the position measurement-based DOb is employed, the generalised Bode's integral equation is as follows:

$$\int_{-\pi}^{\pi} \ln\left|S_P^i\left(e^{j\omega T_s}\right)\right| d\omega T_s = -2\pi \ln\left|1+\psi_P(z)\right| = 0 \quad (18)$$

where $\psi_P(z) = \lim_{z\to\infty} L_P^i(z) = 0$ [31].

Equations (16) and (18) show that as the robustness against disturbances is improved at low frequencies, larger sensitivity peaks appear at middle/high frequencies to hold the Bode's integral equation. When the nominal plant parameters and the bandwidth of the position measurement-based DOb are not properly tuned, *waterbed effect* occurs as illustrated in Fig. 7. This makes the robust motion controller more sensitive to high-frequency disturbances such as noise and degrades the robust stability.

Similar to the velocity measurement-based DOb, the phase margin and robustness (i.e., $\alpha$ and $g_{DOb}$) of the position measurement-based DOb are limited by the stability and *waterbed effect*. However, the design constraints of the position measurement-based DOb are stricter than that of the velocity measurement-based DOb. To relax the design constraints of $\alpha$ and $g_{DOb}$, the sampling time of the motion control system should be similarly decreased.

## IV. DISCRETE-TIME ANALYSIS AND SYNTHESIS OF THE DOb-BASED ROBUST POSITION CONTROL SYSTEM

Figure 8 illustrates the block diagram of the conventional DOb-based robust position control system in continuous-time

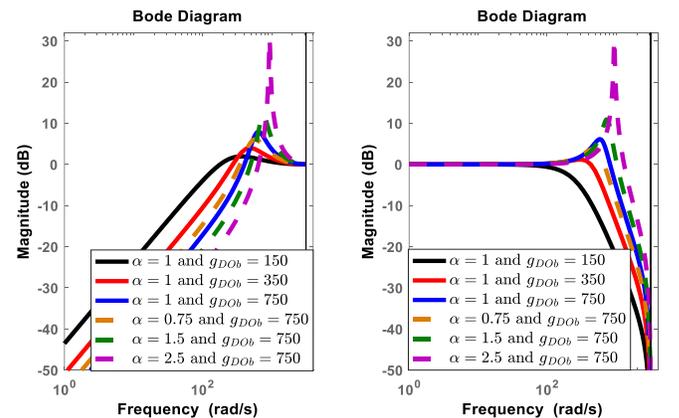

a) Sensitivity function; b) Complementary sensitivity function.
Fig. 7: Frequency responses of $S_P^i(z)$ and $T_P^i(z)$ when $g_v$=750 rad/s and $T_s$=1ms.

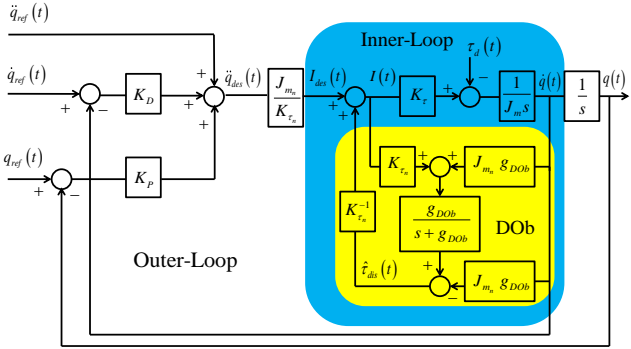

Fig. 8: Block diagram of the DOb-based robust position control system.

domain. In this figure, $K_P$ and $K_D$ are the control gains of the outer-loop performance controller, and $q_{ref}, \dot{q}_{ref}$ and $\ddot{q}_{ref}$ are the position, velocity and acceleration references, respectively. The other parameters are same as defined earlier. The DOb-based robust motion control system has a 2-DOF control structure, i.e., the robustness and performance of the control system can be separately adjusted by tuning the DOb and the performance controller in the inner- and outer- loop, respectively. It is one of the most widely used robust motion control techniques due to its simplicity and efficacy. The reader is referred to [7] for further details on the DOb-based robust control systems in continuous-time domain. In this section, the stability and robustness of the position control system are analysed by considering the acceleration, velocity and position measurement-based DObs in discrete-time domain.

It is noted that various outer-loop performance controllers can be synthesised by using either different discrete-time algorithms (e.g., derivative controller can be implemented by using Backward Euler or Implicit Adams methods when only position measurement is used) or feedbacking velocity and acceleration measurements as well as position measurement. For the sake of brevity and clarity of analysis, the performance controller is synthesised by using position measurement and Backward Euler method as follows:

$$C(z) = K_P + K_D \frac{z-1}{T_s z} \quad (19)$$

Figure 9 illustrates the block diagram of the DOb-based robust position control system and its equivalent block diagram in discrete-time domain. In this figure, * is A, V and P when the acceleration, velocity and position measurement-based DObs are employed in the inner-loop, respectively; and $\tilde{G}_A(z) = G_P(z), \tilde{G}_V(z) = G_P(z)G_V^{-1}(z)$ and $\tilde{G}_P(z) = 1$. $G_V(z)$ and $G_P(z)$ are given in Eqs. (11) and (16), respectively.

The outer-loop's sensitivity and complementary sensitivity functions can be directly derived from Fig. 9b as follows:

$$S_*^o(z) = \frac{1}{1 + L_*^o(z)} \quad (20)$$

$$T_*^o(z) = \frac{L_*^o(z)}{1 + L_*^o(z)} \quad (21)$$

where $L_*^o(z) = C(z)C_*^i(z)G_P(z)$, and $C_*^i(z)$ is given in Eq. (7), Eq. (11) and Eq. (16) when * is A, V and P, respectively.

Equation (20) shows that not only the DOb but also the outer-loop performance controller $C(z)$ can contribute to the robustness of the position control system. In addition, Eq. (21)

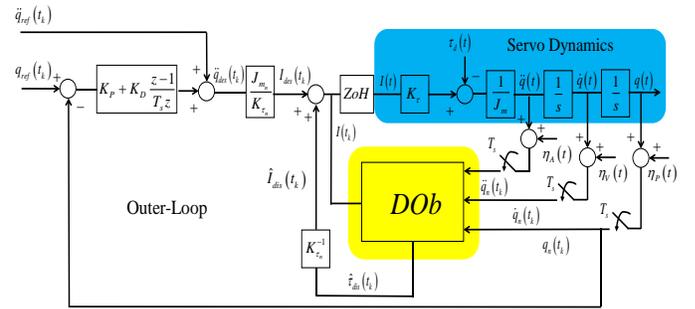

a) DOb-based robust position control system in discrete-time.

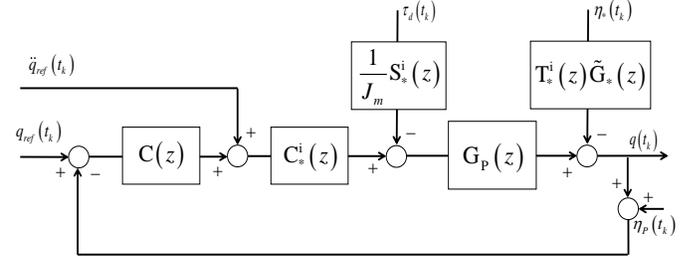

b) Equivalent block diagram of the DOb-based robust position control system.
Fig. 9: DOb-based robust position control system when Eq. (19) is used in the design of the outer-loop performance controller in discrete-time.

shows that the dynamics of the DOb along with $C(z)$ changes the position control performance. The bandwidth of the inner-loop robust controller is generally tuned higher than that of the outer-loop performance controller so that the influence of the dynamics of disturbance estimation on the position control performance is suppressed in practice [6].

Let us consider the robustness of the DOb-based position control system by using the generalised Bode Integral Theorem.

$$\int_{-\pi}^{\pi} \ln\left|S_*^o\left(e^{j\omega T_s}\right)\right| d\omega T_s = 2\pi \left( \ln \sum_{k=1}^{2} \left|p_{u_k}^o\right| - \ln\left|1 + \lim_{z \to \infty} L_*^o(z)\right| \right) = 2\pi \ln \sum_{i=k}^{2} \left|p_{u_k}^o\right| \quad (22)$$

where $p_{u_k}^o$ represents the $k^{th}$ unstable discrete pole of $L_*^o(z)$ [31].

All discrete poles of $L_A^i(z)$ are stable; however, $L_V^i(z)$ and $L_P^i(z)$ may have unstable discrete pole(s) when the design parameters of the DOb are not properly tuned, i.e., $\alpha g_{DOb} > 2/T_s$ (see Section III). Therefore, the robustness constraint of the digital position controller becomes stricter when it is implemented by using velocity and position measurement-based DObs. It is noted that all discrete poles of the open-loop transfer functions are stable when the DOb is synthesised by employing the proposed design constraints in Section III.

Let us first consider the stability. The root-loci of the robust position control systems are plotted with respect to $\alpha$ in Fig. 10. This figure shows that lower values of $\alpha$ degrade the stability of the robust position controllers. As $\alpha$ is increased within a certain range, the phase-lead compensator $C_*^i(z)$ improves the stability. The robust position control system is stable for all high-values of $\alpha$ when it is synthesised by using the acceleration measurement-based DOb. However, the discrete poles of the motion control systems move towards out of the unit circle (i.e., the stability of the robust position controllers deteriorates) as $\alpha$ is increased when the velocity and position measurement-based DObs are employed in the inner-loop.

Not only $\alpha$ but also the bandwidth of the DOb can be used to adjust the stability of the robust position control system. As $g_{DOb}$ increases, the stability improves by compensating the

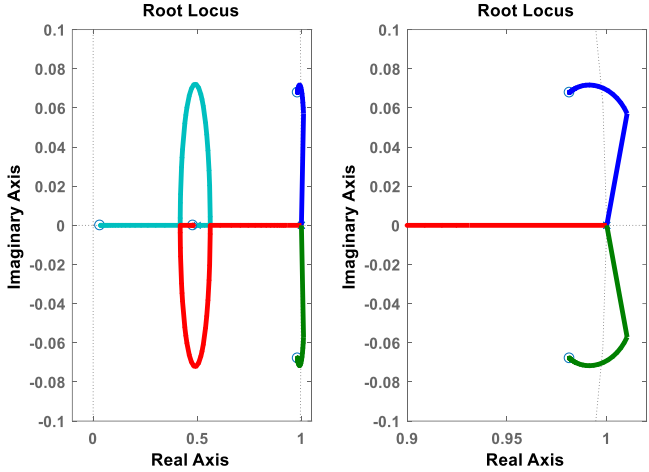

a) Acceleration measurement-based DOb. Left figure illustrates the asymptotic behaviour of the root locus and right figure illustrates its local behaviour between 0.9 to 1.05 on the real-axis and -0.1 to 0.1 on the imaginary-axis.

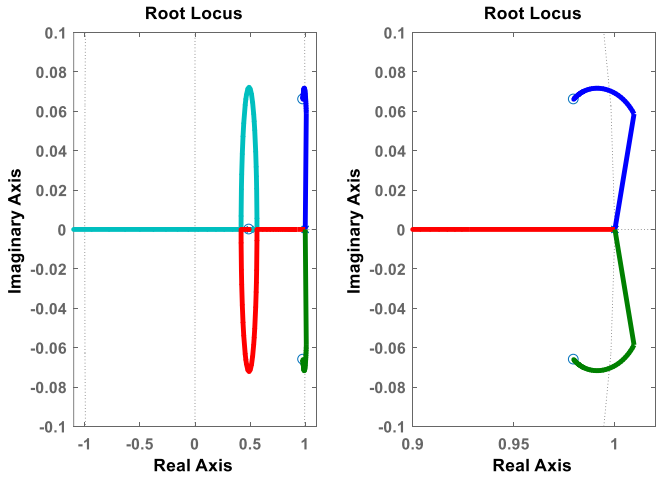

b) Velocity measurement-based DOb. Left figure illustrates the asymptotic behaviour of the root locus and right figure illustrates its local behaviour between 0.9 to 1.05 on the real-axis and -0.1 to 0.1 on the imaginary-axis.

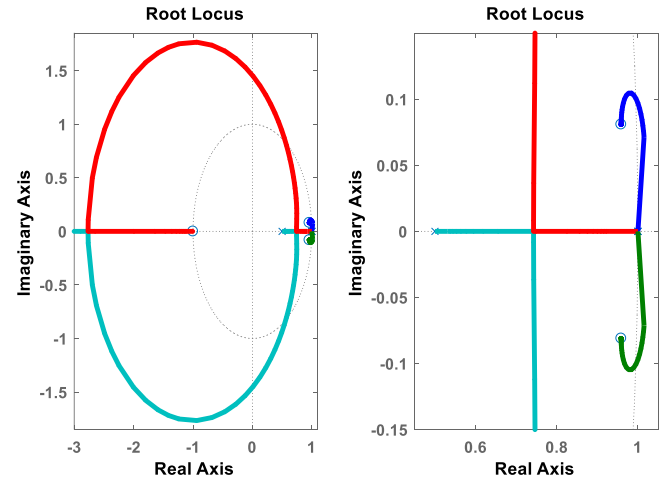

c) Position measurement-based DOb. Left figure illustrates the asymptotic behaviour of the root locus and right figure illustrates its local behaviour between 0.45 to 1.05 on the real-axis and -0.15 to 0.15 on the imaginary-axis.
Fig. 10: Root-loci of the robust position control system with respect to α when the DOb is implemented by using acceleration, velocity and position measurements. The parameters of the simulations are $J_m = 0.003$, $K_\tau = 0.25$, $g_{DOb} = 500$, $g_v = 1000$, $K_P = 5000$, $K_D = 25$, and $T_s = 1$ms.

parametric uncertainties due to the variations of inertia and thrust coefficient, i.e., $\alpha$. However, the bandwidth of the DOb may also have an upper-bound due to the stability constraints derived in Section III. The robust position control system may become unstable as the bandwidth of the DOb is increased. It is illustrated for the velocity measurement-based DOb in Fig. 11. To eliminate the stability constraint on the bandwidth of the observer, the acceleration-measurement-based DOb should be employed in the inner-loop.

Let us now consider the robustness. Figure 12 illustrates the frequency responses of the sensitivity and complementary sensitivity functions of the inner- and outer- loop when the acceleration, velocity and position measurement-based DObs are employed in the inner-loop. The frequency responses of the exogenous noise and disturbance inputs are also illustrated in this figure.

When the DOb is implemented in the inner-loop, the robustness against disturbances of the position control system can be significantly improved. For example, the peaks of the outer-loop sensitivity functions can be reduced and the overall disturbance suppression can be increased by $S_*^i(z) \times S_*^o(z)$ in the low frequency range (see left column in Fig. 12). However, if the design parameters of the DOb are not properly tuned, then the sensitivity function of the position control system may have a large peak value at high frequencies (e.g., see left column in Fig. 12c).

Since the dynamics of the inner-loop is tuned faster than that of the outer-loop, the noise response of the position control system highly depends on the dynamics of the DOb at high frequencies (see right column in Fig. 12). However, although the dynamics of the DOb is properly tuned in the inner-loop so that the peak of $T_*^i(z)$ is low, middle/high frequency disturbances may still be excited by the outer-loop performance controller (e.g., see right columns in Fig. 12a and Fig. 12b). As shown in Fig. 12c, not only the DOb but also the outer-loop performance controller should be properly tuned to suppress the peak of the complementary sensitivity function, i.e., noise-sensitivity. Figure 13 illustrates how the peak of the complementary sensitivity function can be reduced by properly tuning the outer-loop performance controller when the velocity measurement-based DOb is employed in the inner-loop.

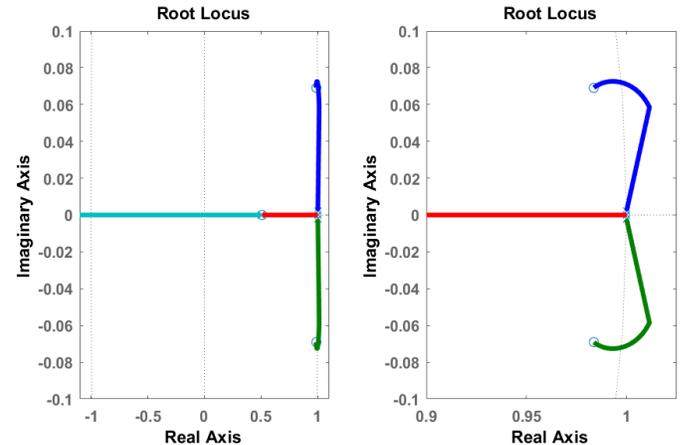

Fig. 11: Root-locus of the robust position control system with respect to $g_{DOb}$ when the DOb is implemented by using velocity measurement. Left figure illustrates the asymptotic behaviour of the root locus and right figure illustrates its local behaviour between 0.9 to 1.05 on the real-axis and -0.1 to 0.1 on the imaginary-axis. The simulation parameters are $J_m = 0.003$, $J_{mn} = J_m \times 10^{-3}$, $K_{\tau n} = K_\tau = 0.25$, (i.e., α = 0.001), $K_P = 5000$, $K_D = 25$, and $T_s = 1$ms.

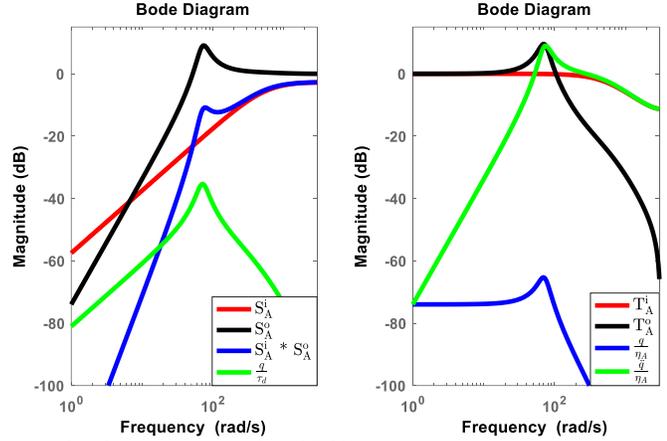

a) Acceleration measurement-based DOb.

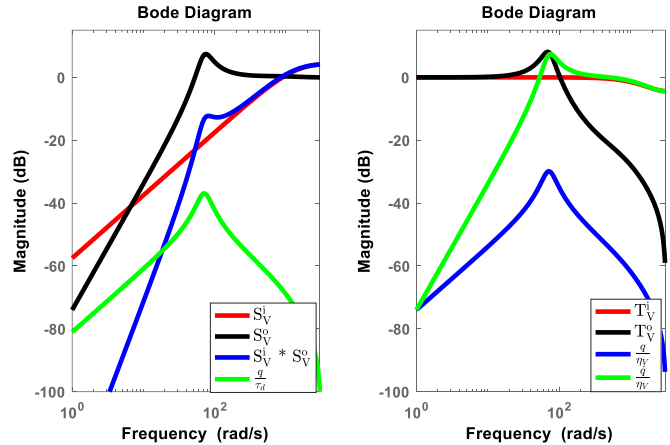

b) Velocity measurement-based DOb.

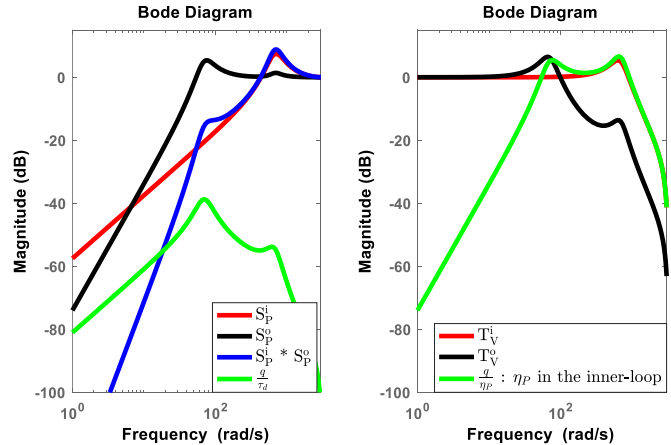

c) Position measurement-based DOb.

Fig. 12: Frequency responses of the sensitivity and complementary sensitivity functions and that of exogenous noise and disturbance inputs. The parameters of the simulations are $J_m = 0.003$, $J_{mn} = J_m \times 1.5$, $K_{\tau n} = K_\tau = 0.25$ (i.e., $\alpha = 1.5$), $g_{DOb} = 500$, $g_v = 1000$, $K_P = 5000$, $K_D = 25$, and $T_s = 1$ms.

## V. EXPERIMENTS

To validate the proposed analysis and synthesis methods, we applied the DOb-based robust position controllers to the servo system shown in Fig. 14. The experimental setup was built by using Maxon RE25 DC motor, ESCON 50/5 motor driver, Renishaw RGH24 linear encoder to measure the position of the linear guide, DCT22 tachometer to measure the angular speed of the motor shaft and Memsic CXL04GP3 accelerometer to measure the acceleration of the linear guide. A PC with a Linux operating system was employed to perform the real-time motion control experiments.

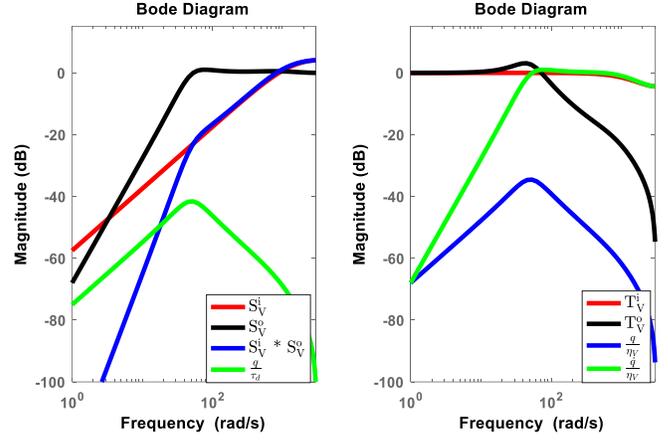

Fig. 13: Frequency responses of the sensitivity and complementary sensitivity functions and that of exogenous noise and disturbance inputs. The parameters of the simulations are $J_m = 0.003$, $J_{mn} = J_m \times 1.5$, $K_{\tau n} = K_\tau = 0.25$, $g_{DOb} = 500$, $g_v = 1000$, $K_P = 2500$, $K_D = 50$, and $T_s = 1$ms.

Let us start with regulation control experiments and validate the performance of the robust position controllers synthesised by employing the acceleration, velocity and position measurement-based DObs in the inner-loop. The robust motion controllers were tuned by using the exact plant parameters, i.e., $\alpha = 1$. Another servo system was attached to the linear guide and applied 4N external disturbance force between 4 and 6 seconds. A force sensor was placed between the servo systems to measure the contact force. Figure 15 illustrates the robust position control experiments. It is clear from this figure that the robust position controller can precisely suppress the internal and external disturbances and follow the step reference input when it is synthesised by employing the acceleration, velocity and position measurement-based DObs. The difference between the applied force measured by the force sensor and the disturbances estimated by the DOb is due to the internal disturbances such as friction and backlash of the speed reducer.

Let us now consider the stability of the robust motion controller. The regulation control experiments were conducted by using different values of the nominal inertia (i.e., $\alpha \neq 1$) when the servo system was not disturbed by an external load. Figure 16a shows that the robust motion controller is stable for

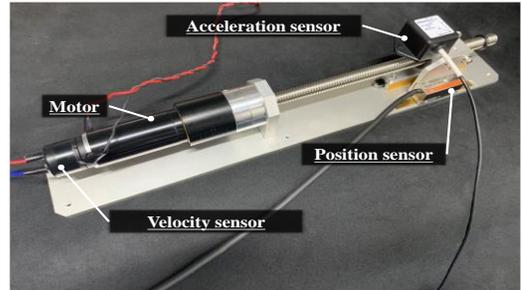

Fig. 14: Experimental setup.

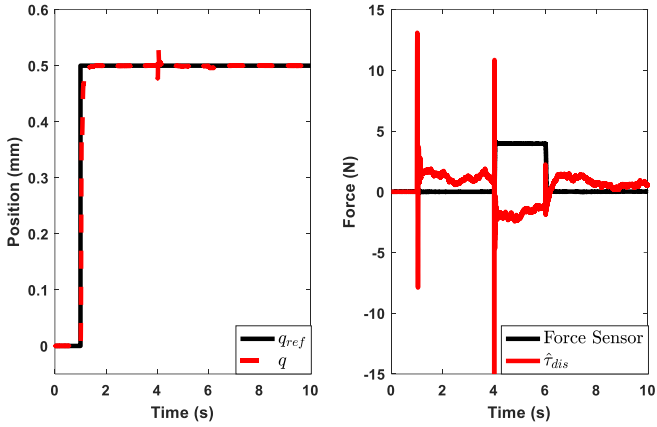

a) Acceleration measurement-based DOb.

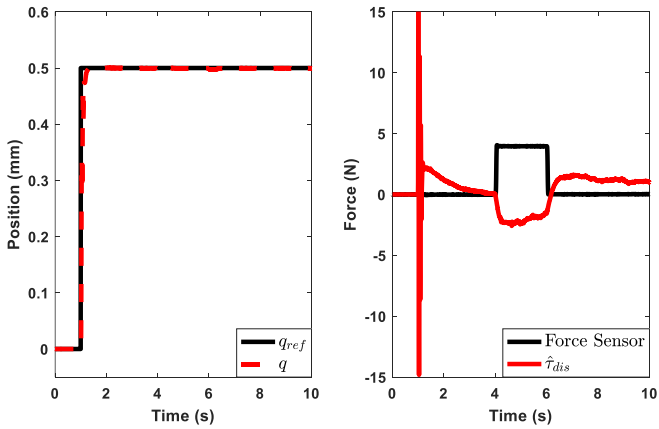

b) Velocity measurement-based DOb.

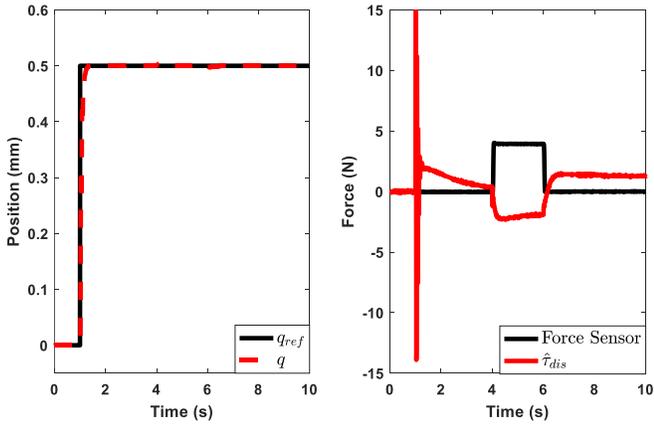

c) Position measurement-based DOb.

Fig. 15: Regulation control experiment when 4N external disturbance was applied by using another servo system. In this experiment, $\alpha = 1$, $g_{DOb} = 1000$, $g_v = 2000$, $K_P = 4000$, $K_D = 200$, and $T_s = 0.5$ms.

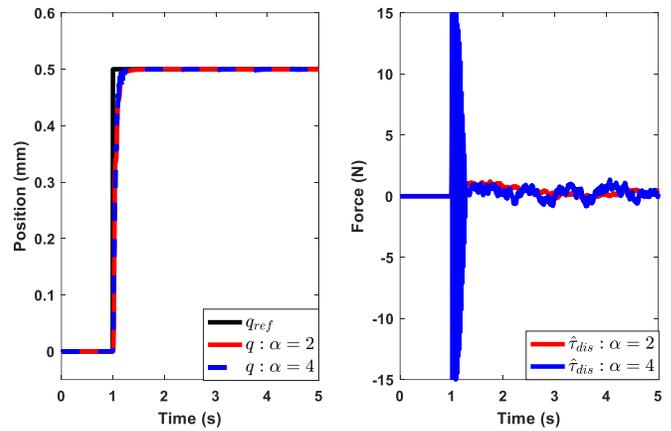

a) Acceleration measurement-based DOb.

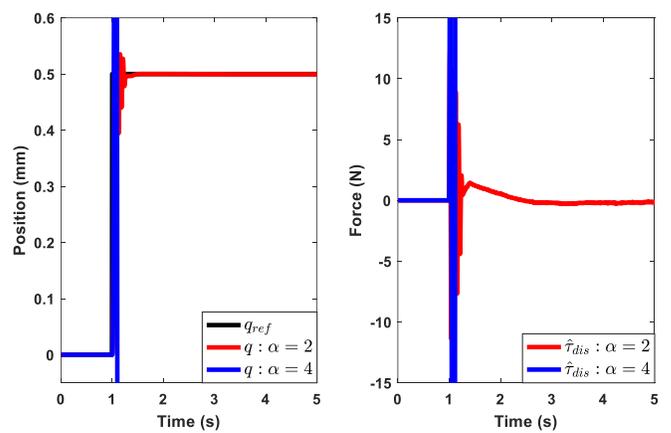

b) Velocity measurement-based DOb.

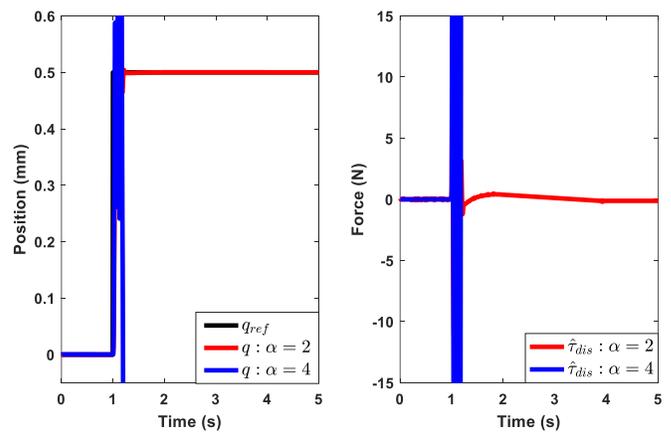

c) Position measurement-based DOb.

Fig. 16: Regulation control experiment when different nominal inertiae were used in the design of the DOb. In this experiment, $g_{DOb} = 1000$, $g_v = 2000$, $K_P = 4000$, $K_D = 200$, and $T_s = 0.5$ms.

all values of $\alpha$ when the acceleration measurement-based DOb is employed in the inner-loop. However, the estimation of disturbances and robust control signal become more sensitive to noise for higher values of $\alpha$. Figures 16b and 16c show that the stability deteriorates as $\alpha$ is increased when the robust motion controller is synthesised by using the velocity and position measurement-based DObs. The robust position controllers were unstable when $\alpha \geq 4$.

Last, let us present trajectory tracking control experiments. The robust position controllers were synthesised by tuning $\alpha = 1$ and the servo system was not disturbed by an external load. Figure 17 shows that the DOb-based robust position controllers can provide high performance in trajectory tracking control. In this experiment, a sinusoidal trajectory reference was precisely tracked by suppressing internal disturbances. The noise-sensitivity of disturbance estimation was similar when the DOb was synthesised by employing acceleration, velocity and position measurements.

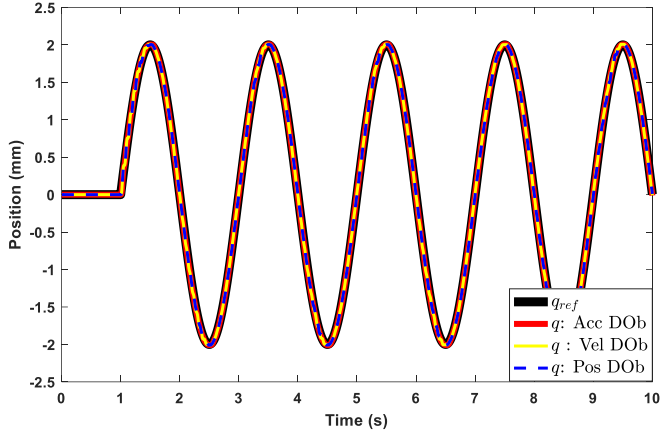

a) Trajectory tracking control.

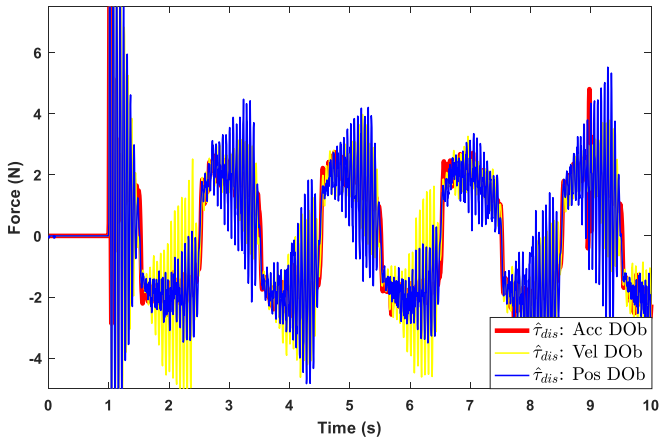

b) Estimation of disturbances.

Fig. 17: Trajectory tracking control experiment when no external load was applied to the servo system. In this experiment, $\alpha = 1$, $g_{DOb} = 1000$, $g_v = 2000$, $K_P = 4000$, $K_D = 200$, and $T_s = 0.5$ms.

## VI. CONCLUSION

This paper presents new stability and robustness analyses for the DOb-based robust motion control systems in discrete-time domain. It is theoretically and experimentally shown that the robust stability and performance of the digital motion controller significantly change when acceleration, velocity and position measurements are employed in the DOb synthesis. The acceleration measurement-based DOb provides good robust stability and performance for all values of the design parameters of $\alpha$ and $g_{DOb}$. The plant-model mismatch (i.e., phase margin) and the bandwidth of the DOb (i.e., robustness) are constrained by only the noise of acceleration measurement. However, in addition to the trade-off between the robustness and noise-sensitivity, the design parameters of the velocity and position measurement-based DObs are constrained by *waterbed effect*. The robust stability and performance of the digital motion controller may significantly deteriorate if the design parameters of the velocity and position measurement-based DOb are not properly tuned. By employing the proposed stability and robustness constraints, one can systematically synthesise a high-performance DOb-based digital robust motion controller in practice. For example, the nominal design parameters and the bandwidth of the DOb can be automatically tuned by adaptive robust digital motion controllers [32]. Moreover, advanced motion controllers can be synthesised by shaping the sensitivity and complementary sensitivity transfer functions given in Eqs. (7), (11), (16), (20) and (21) in discrete-time.

Sections II and III show that the design constraints on the nominal plant model (i.e., $\alpha$) and the bandwidth of the DOb (i.e., $g_{DOb}$) implemented by computers cannot be derived in continuous-time domain. Moreover, continuous-time analysis methods cannot explain some dynamic responses of the DOb-based digital motion control systems. This may cause severe robust stability and performance issues, e.g., the robust motion controller may unexpectedly become sensitive to internal and external disturbances. Thus, this paper recommends discrete-time analysis and synthesis methods for tuning the design parameters of the DOb-based digital motion control systems.

## VII. ACKNOWLEDGEMENT

This work was supported in part by the Ministry for Education, Science and Youth of Sarajevo Canton and by the Ministry of Civil Affairs of Bosnia and Herzegovina.

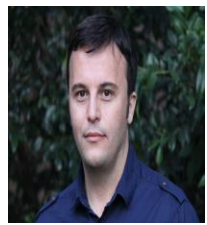
**Emre Sariyildiz** (S'11–M'16) received the PhD degrees in Integrated Design Engineering from Keio University, Yokohama, Japan, in 2014 and in Control and Automation Engineering from Istanbul Technical University, Istanbul, Turkey, in 2016. He was as a Research Fellow in the Department of Biomedical Engineering at National University of Singapore, Singapore, between 2014 and 2017. Since April 2017, he has been a Lecturer with the School of Mechanical, Materials, Mechatronic, and Biomedical Engineering, University of Wollongong, Wollongong, NSW, Australia. His research interests include control theory, mechatronics, robotics and motion control.

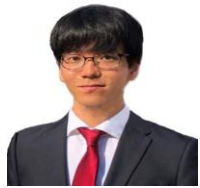
**Satoshi Hangai** (S'18-M'20) received the B.E. degree in system design engineering and the M.E. degree in integrated design engineering from Keio University, Yokohama, Japan, in 2018 and 2020, respectively. His research interests include haptics and motion control.

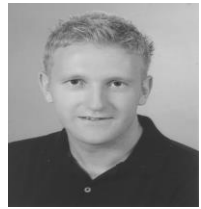
**Tarik Uzunovic** (S'09-M'16-SM'19) received the B.Eng. and M.Eng. degrees in electrical engineering from the University of Sarajevo, Sarajevo, Bosnia and Herzegovina, and Ph.D. degree in mechatronics from Sabanci University, Istanbul, Turkey, in 2008, 2010, and 2015, respectively. He is an Associate Professor with the Department of Automatic Control and Electronics, Faculty of Electrical Engineering, University of Sarajevo, Sarajevo, Bosnia and Herzegovina. His research interests include control theory, motion control, robotics, and mechatronics.

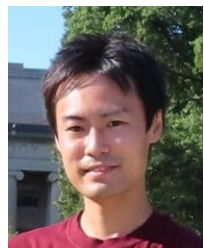
**Takahiro Nozaki** (S'11-M '13) received his B.E., M.E., and Ph.D. degrees from Keio University, Yokohama, Japan, in 2010, 2012, and 2014, respectively. In 2014, he joined Yokohama National University, Yokohama, Japan, as a Research Associate. In 2015, he joined the Keio University, where he is currently an Assistant Professor. He is with Massachusetts Institute of Technology, Massachusetts, United States, as a Visiting Scientist. He was selected as a winner of the IEEE Industrial Electronics Society Under 35 Innovators Contest in 2019.

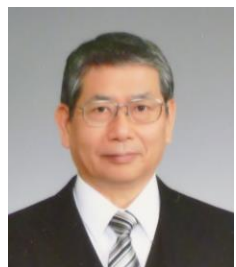
**Kouhei Ohnishi** (S'78, M'80, SM'00, F'01, LF'17) received the B.E., M.E., and Ph.D in electrical engineering from the University of Tokyo in 1975, 1977 and 1980. Since 1980, he has been with Keio University.

He served as a President of the IEEE Industrial Electronics Society in 2008 and 2009 as well as a President of IEEJ in 2015 and 2016. Since 2016, he has been also with Kanagawa Institute of Industrial Science and Technology at Kawasaki, Japan.

He received the Medal of Honor with Purple Ribbon from His Majesty the Emperor in 2016.